\documentclass[sigconf,screen]{acmart}

\usepackage{booktabs}
\usepackage{listings}
\usepackage{graphicx}
\usepackage{algorithm}
\usepackage{algorithmic}
\usepackage{hyperref}
\usepackage{xcolor}
\usepackage{breakcites}
\usepackage{natbib}
\usepackage{soul}

\setcitestyle{numbers}
\renewcommand\footnotetextcopyrightpermission[1]{}
\acmConference[EASE 2025]{The 29th International Conference on Evaluation and Assessment in Software Engineering}{17--20 June, 2025}{Istanbul, Turkey}
\acmYear{2025}
\copyrightyear{2025}

\title{Towards Requirements Engineering for RAG Systems}

\author{Tor Sporsem}
\email{tor.sporsem@sintef.no}
\orcid{0000-0002-5230-7480}
\affiliation{%
  \institution{SINTEF \& NTNU}
  \city{Trondheim}
  \state{}
  \country{Norway}
}

\author{Rasmus Ulfsnes}
\email{rasmus.ulfsnes@sintef.no}
\orcid{0000-0002-4966-8242}
\affiliation{%
  \institution{SINTEF \& NTNU}
  \city{Trondheim}
  \state{}
  \country{Norway}
}


\begin{document}

\begin{abstract}
This short paper explores how a maritime company develops and integrates large-language models (LLM). Specifically by looking at the requirements engineering for Retrieval Augmented Generation (RAG) systems in expert settings. Through a case study at a maritime service provider, we demonstrate how data scientists face a fundamental tension between user expectations of AI perfection and the correctness of the generated outputs. Our findings reveal that data scientists must identify context-specific "retrieval requirements" through iterative experimentation together with users because they are the ones who can determine correctness. We present an empirical process model describing how data scientists practically elicited these "retrieval requirements" and managed system limitations. This work advances software engineering knowledge by providing insights into the specialized requirements engineering processes for implementing RAG systems in complex domain-specific applications.
\end{abstract}

\maketitle

\section{Introduction}

Kat, a maritime engineer at Marcomp, has received a question from a ship's captain. Her job is to help Marcomp’s customers understand and apply international regulations that ships must follow. This time, instead of drafting the answer herself, she uses a recently provided Large Language Model (LLM). She clicks the “generate reply” button, and within seconds, a response appears on her screen. Smiling, she says: 
\begin{quote}
    "Incredible right? It has searched through all the 500,000 previous answers we have given, found similar ones and written a new answer based on all the old ones. Now, let's check if we can trust it"
\end{quote}
She reads through the generated answer and notices this would have been a great answer a few years ago. However, new rules have been implemented, making the generated answer incorrect.
\begin{quote}
    "Well, I know this rule is quite new and there are probably no previous answers on similar questions after the new rule came into effect."
\end{quote}
Kat deletes the generated text and writes up her own answer. 

This highlights a fundamental problem of "plugging in" an LLM into an organization's local data (knowledge base) -- using a retrieval augmented generation (RAG) system. The knowledge base, which contains 500,000 answers to questions about ships from the past 15 years, holds no answers where the new rule has been in effect. When the LLM is instructed to predict its answer based on previous answers, it cannot identify which are outdated and consequently generates incorrect answers.

There are already studies describing how software engineering (see e.g. \citet{wan_how_2021, kalinowski_naming_2024}) and requirements engineering (see e.g. \citet{kim_data_2018, amershi_software_2019}) change when developing AI systems. However, RAG is gaining in popularity because of its pragmatic way of integrating LLM into organizations' databases without having to train their own \cite{barnett_seven_2024}. With this as a backdrop, we ask the following research question: \emph{How do developers elicit requirements for RAG systems?}   

In this short paper, we present preliminary findings from a case study that shows how a team of developers and data scientists implemented RAG in a maritime company. 
\section{Background}

\subsection{Retrieval augmented generation}

While LLMs excels at various tasks, including software engineering \cite{ulfsnes_transforming_2024} and across experimental applications in different fields \cite{bubeck_sparks_2023}, there remains a gap in their training data that leads to inconsistent results \cite{dellacqua_navigating_2023}. Re-training an LLM is impractical for small applications due to the extensive time required (months) and smaller organizations' resource constraints \cite{barnett_seven_2024}. RAG proposed by Facebook AI Research in 2020, offers a potential solution that avoids retraining costs while leveraging external data \cite{lewis_retrieval-augmented_2020}. The RAG process involves four key steps:

\begin{enumerate}
    \item Indexing - Converting existing knowledge into a machine-understandable knowledge base, a vector database with embeddings, that are semantic representations of the words, and their relation to other words in the documents.
    \item Retrieval - Finding relevant stored information in the knowledge base when presented with an external query.
    \item Augmentation - Combining the incoming query with the retrieved information from the knowledge base.
    \item Generation - Using the augmented information to produce an LLM output.
\end{enumerate}

Current literature on RAG primarily focuses on technical aspects, such as selecting appropriate indexing, embedding strategies, and retrieval technologies \citep{veturi_rag_2024} (e.g., ScaNN). Some studies evaluate RAG efficiency across use cases like QA, text generation, summarization, and SW development and maintenance \citep{arslan_survey_2024}. Another line of work examines technical implementation challenges, identifying failure points related to missing content, lack of relevant docs, or incomplete answers \citep{barnett_seven_2024}. This research also highlights the need for ML skills. While various ENG strategies and optimization approaches are being explored, literature on RE and AI development indicates that in expert domains, close user involvement is often necessary \citep{ishikawa_how_2019, waardenburg_coexistence_2022}.

\subsection{Requirements Engineering for AI}
While research on the development of RAG is still in its infancy, research on how to develop ML systems has been extensively researched in the last few years \cite{ahmad_requirements_2023}. Similar to ML systems, when developing RAG systems a fundamental requirement is that the knowledge base is rich and preferably complete \cite{barnett_seven_2024}. This means that the LLM should be able to find relevant information in the knowledge base on which to build its generation, and reduce hallucination. Research on machine learning has shown that a knowledge base with large amounts of data and a high degree of diversity increases the chance that the ML system can handle a query \citep{vogelsang_requirements_2019}. For example, if an ML system is designed for doctors, the knowledge base needs to include a diverse variety of diagnoses and as many characteristics of each diagnosis as possible \cite{lebovitz_is_2021}. A lack of diversity means a risk of missing diagnoses, which can be critical \cite{vogelsang_requirements_2019}. The number of examples in the knowledge base is thus not as significant as the diversity in examples \citep{vogelsang_requirements_2019}. Moreover, this principle applies across medical \cite{lebovitz_is_2021}, human resources \cite{van_den_broek_when_2021}, or technical domains \cite{amershi_software_2019}. The quality and breadth of information determine how effectively the system handles common and specialized queries. 

Another fundamental challenge when engineering AI systems lies in establishing clear correctness criteria for system outputs \citep{ishikawa_how_2019}. This difficulty stems from the inherent ambiguity in determining what constitutes a "correct" prediction, as correctness can manifest in various forms and interpretations \citep{lipton_mythos_2018}. Unlike systems with binary or limited output possibilities, LLM outputs rarely have a single definitive correct answer \cite{ulfsnes_generation_2024}. Even when addressing factual questions with a RAG system with a trusted knowledge base, multiple valid expressions of the same information exist.
Additionally, the notion of correctness in these systems may evolve as real-world conditions change over time. This variability in what constitutes correctness presents significant challenges for developing evaluation frameworks and benchmarks and performance benchmarks for LLMs \citep{chen_benchmarking_2024}. This leads to the question of who can judge if an output of a RAG system is correct or not and if an output is "correct enough" to be of value. 

Research already demonstrates that eliciting requirements for AI systems introduces unique challenges \citep{kalinowski_naming_2024, vogelsang_requirements_2019}. Given the black box nature of most AI models, requirements engineers struggle to specify precise requirements, often producing specifications considered 'too high-level' or 'vague' \citep{ahmad_requirements_2023}. Rather than focusing on precise initial requirements, it is recommended to first hypothesize possible outcomes from available data \citep{wan_how_2021}, then refining requirements through experimentation \citep{ishikawa_how_2019, giray_software_2021}. However, we still lack empirical industry cases documenting how developers and data scientists elicit requirements for RAG systems.

\section{Methods}

\begin{table}
  \caption{Collected data}
  \label{tab:data}
  \begin{tabular}{ll}
    \toprule
    Data source&Type of data\\
    \midrule
    Idea Workshop & 7 hours observation during workshop \\
    Data Architect & 7 hours observation \& 2 interviews  \\
    AI Solution Engineer & 3 interviews \\
    8 Case handlers (users) & 28 hours of observation \& 4 interviews \\
    
  \bottomrule
\end{tabular}
\end{table}

\textit{Study design \& case description.} The research literature on SE and RE benefits from case studies that provide insights into how these processes unfold in the industry. Case studies are particularly suitable for evidence-based recommendations for practitioners \cite{stol_teaching_2024}. Following this, we embarked on a case study\cite{runeson_guidelines_2008} following Marcomp's AI software development and present initial findings from one project. 

Marcomp is a major maritime sector service provider with approximately 3,700 employees worldwide. Their primary business involves verifying vessels' compliance with regulations; successful verification results in certificates that ships need for obtaining marine insurance and operating internationally. As a customer of Marcomp, shipping companies also have access to a 24/7 helpdesk where they can get support from Marcomp's top engineers on everything from technical questions to regulatory questions. Marcomp receives hundreds of questions every day from customers around the world and has five offices in different time zones to operate the help desk, with approximately 60 case handlers. Marcomp maintains an internal development department that creates software for Marcomp employees and external customers. 
%Development is structured as a project portfolio with a matrix organization, allowing development resources to be allocated to projects based on need. The list of desired projects significantly exceeds capacity, resulting in constant demand for software developers and data scientists.

\textit{Data collection and analysis.} To completely understand the case, we interviewed and observed developers and users (case handlers). Interviews are a valuable source of information as they allow in-depth conversation with experts. Still, it is only one of many potential methods to collect data in field studies \cite{lethbridge_studying_2005}. Interviews fall in a category of methods that \citet{lethbridge_studying_2005} have labeled inquisitive techniques, in that a researcher must actively engage with interviewees to get information from them. A second category, "observational techniques," includes watching professionals at work. Each approach has strengths and limitations; interview data may be less accurate than observational data, while observation may trigger the Hawthorne effect, where people modify their behavior when they know they're being watched \cite{lethbridge_studying_2005}. 

In the summer of 2023, we observed an idea development workshop with managers, case handlers, and data scientists who decided to explore if LLMs could be useful for case handlers in answering customer questions. Then, a few months later, in late 2023, we observed four case handlers to understand their work process and software use. We spent three full days with each of them, observing how they answered customer questions and observed their meetings and lunch discussions. Additionally, we interviewed four other case handlers. In early 2024, we interviewed two data scientists (a Data Architect and an AI Solution Engineer) to understand their approach and working methods. Again, in late 2024, we revisited the four case handlers and data scientists to observe and interview them after launching the RAG system. All data is summarized in table \ref{tab:data}.

For our preliminary data analysis presented in this paper, we started with a grounded theory approach \cite{stol_grounded_2016}, combined with Seaman's \citep{seaman_qualitative_1999} guidelines for open-ended coding and memoing. Based on this initial analysis, we noticed different phases emerge from the process of developing the RAG system. We then looked at temporal bracketing strategy from  Langley \cite{langley_strategies_1999} to guide the analysis further. We identified five different phases (summarized in figure \ref{fig:rag process}): 1) Knowledge modeling and experimentation, 2) Retrieval strategy, 3) Retrievable data management, 4) Monitoring and operation, and 5) Expectation management. 

\section{Results}

The release of ChatGPT in November 2022 created tremendous excitement among businesses, with managers impressed by its capabilities. This wave also caught Marcomp. Their senior management was determined to prevent competitors from gaining an edge by adopting this technology first. This led to middle managers rapidly developing ideas for AI implementation, with funding for AI projects becoming readily available. Fortunately for Marcomp, they began building their AI expertise in 2017 by forming a data science team to investigate the growing opportunities in Deep Learning.

\subsection{Knowledge Modeling and experimentation}
\begin{figure}
    \centering
    \includegraphics[width=1\linewidth]{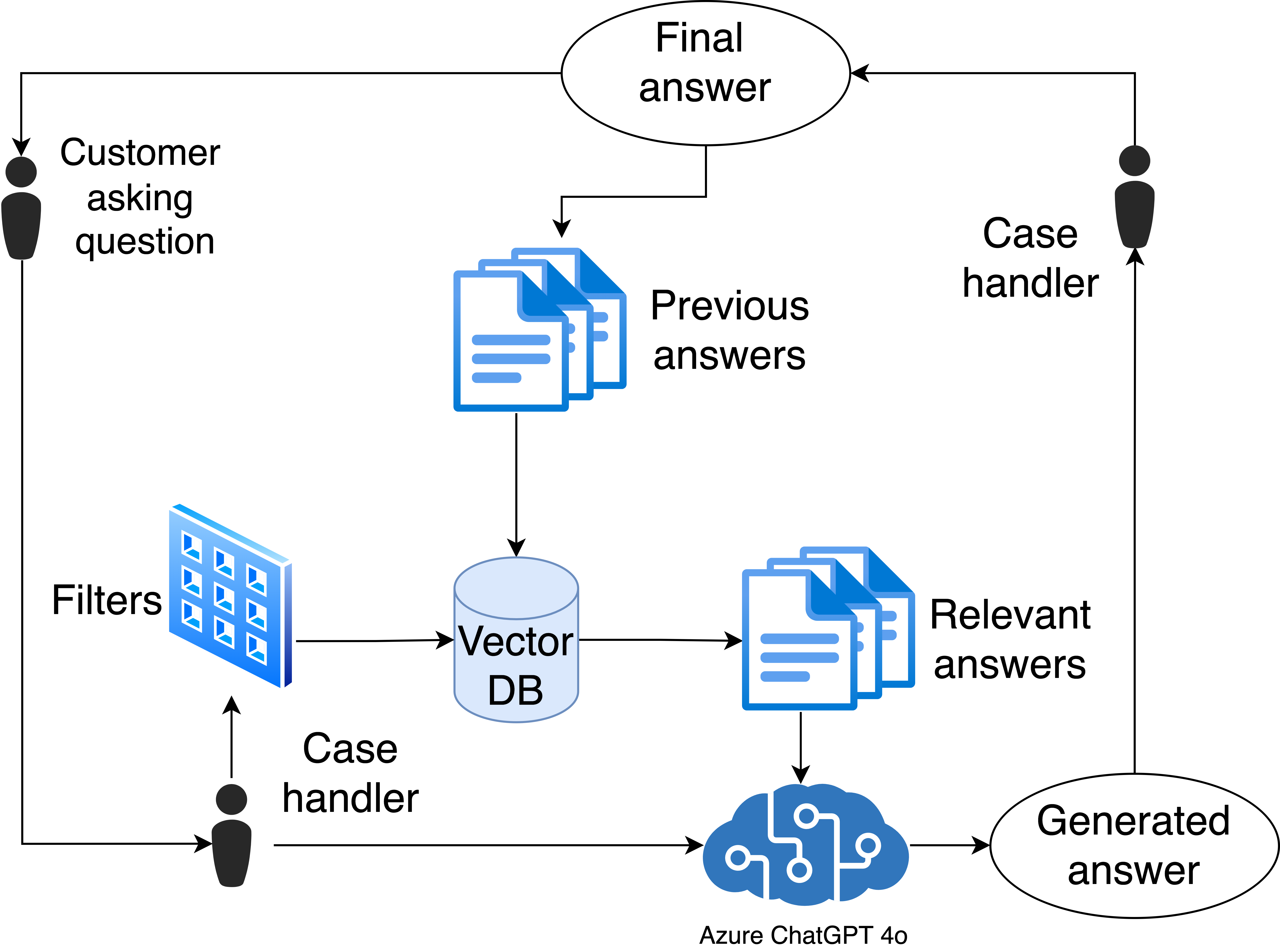}
    \caption{The RAG system as developed by Marcomp. First, a question—e.g., from a ship's captain—is sent to a case handler. The case handler sets filters to help the RAG system retrieve relevant past answers. These, along with the question, form the input for the LLM. The generated answer is then reviewed and often revised by the case handler before sending a final answer back to the captain.}
    \label{fig:rag marcomp}
\end{figure}

During a two-day workshop in 2023, case handlers' managers and data scientists developed the idea that LLMs could extract value from Marcomp's accumulated data. They recognized that their database of 500,000 preserved case handler responses could be a valuable resource to support their current team of 60 case handlers. They assumed that the knowledge needed to answer questions was contained in these 500,000 previous answers and could be utilized by an LLM. They discussed creating a RAG system that combined a GPT-4 large language model with the knowledge base of 500,000 answers (see figure \ref{fig:rag marcomp}). This system was designed to match incoming customer questions with similar previous responses and generate new answers based on these historical examples.

But were 500,000 previous answers enough for the model to generate correct answers (in other words, was the knowledge base suitable)? The data scientists understood that defining the correctness of the model's generated answers would not be straightforward. 'Correct' could mean different things in different contexts. Additionally, the data scientists needed the domain knowledge of case handlers to determine correctness. They decided the best approach was to develop the RAG system and begin experimenting with the case handlers, allowing them to judge the 'correctness' of the answers.
\begin{quote}
    {``This is an experiment; it will be interesting to see if this GPT model can actually generate sensible answers that convince the case handlers.'' -- Heimdal, AI Solution Engineer}
\end{quote}

Their worry with this approach was that an early release of a poorly performing model might disappoint users, who could then dismiss the entire system.
\begin{quote}
``We can keep talking forever, but if we are going to figure out if this can work, then we just need to start.'' -- Heimdal, AI Solution Engineer
\end{quote}
The initial feedback from test users was positive. Junior case handlers were particularly enthusiastic, seeing this as a tool that could compensate for their limited experience and knowledge. They could now benefit from all previous answers created by their senior colleagues. Additionally, since Marcomp operates globally, most case handlers are not native English speakers, and many felt they could now produce replies that better communicated their message in English.

\subsection{Retrieval strategies}

Looking back to the scenario in the introduction, with Kat having to delete the generated answer because a new rule came into effect, one can see a significant data requirement challenge: \textit{changing rules over time}. When a new rule takes effect or an existing rule changes, the database of previous answers lacks examples reflecting these updates. For instance, if a new rule became effective in 2018, all previously stored answers might be inaccurate depending on the change in the rule.

\begin{quote}
    ``The model will make mistakes because it only has examples from the old rules and not the new ones.'' -- Magnus, Data Architecht
\end{quote}

Since they had a fixed set of previous answers, and it would take time for case handlers to generate new responses that followed updated rules, they needed another solution. Their idea was to add an always-updated rulebook to the existing database of answers. However, they discovered that linking specific ships to relevant rules was more complicated than anticipated.

\begin{quote}
    ``I did what turned out to be a silly project at the start where I just took all the rules and ran them through a standard vectorization solution, but it didn't work because then you get that dilemma where you ask questions about passenger ships, it responds with things that apply to container ships. It doesn't distinguish the contexts.'' -- Magnus, Data Architecht
\end{quote}

The rules were too similar for the model to connect them to the specific ship in question correctly. As a result, the data scientists abandoned trying to solve this challenge by adding more data. Instead, they developed what they called a "filtering" function. This meant the case handler could filter out all the previous answers according to a year. So, if a rule took effect in 2018, they could instruct the model to exclude all answers prior to 2018. This meant that in order for case handlers to generate a good answer, they had to support the model in filtering out the irrelevant previous answers. 

\begin{quote}
    ``To accomplish this [applying correct filters], they need many years of experience [as a case handler]. ...  They need to have a complete overview of all the rules to understand which rules apply to the ship in question.'' -- Magnus, Data Architecht
\end{quote}

Unfortunately, rules were not the only contextual challenge. Case handlers quickly discovered that the model struggled with unfamiliar contexts and sometimes created fictional responses.

\begin{quote}
    Sometimes, it hallucinates. It picks up information that is not necessarily relevant for that specific question. -- Karl, Case handler
\end{quote}
A hallucinating LLM indicates a gap in the previous-answers-database. 
\begin{quote}
    ``Every ship is different, even sister ships have differences. This is because different people manage the ships, different management styles, different operational areas, cargoes, and flags [nationalities]. ... So probably it [the RAG system] would take the answer from an already answered question, which is of similar age, similar type of vessel, similar flag, but not necessarily the same management.'' – Emil, Case handler
\end{quote}

Since the contextual factors are numerous and can combine in countless ways, creating almost infinite scenarios. Questions will emerge with combinations of contextual elements not found in previous answers. When these new combinations do not exist in the database, the model does not recognize the context and lacks a framework for handling the question. Case handlers have a significant advantage over the model when dealing with this complexity—they can gather information from beyond computer systems. We observed them calling surveyors who had inspected the vessel to hear what they observed. Or they consulted colleagues in other departments who had dealt with similar cases to get necessary explanations. In this way, case handlers obtained vital information not stored in the database of previous answers or any computer system. Case handlers doubted the model's ability to manage high complexity because they knew it would not access information about all these factors.

To address the context problem, the data scientists enhanced the filtering function to include multiple ship characteristics such as nationality, type of vessel, age, etc. This allowed case handlers to filter previous answers that shared similar contextual factors. When case handlers clicked "generate reply," the system produced text based on these filtered previous answers, increasing the likelihood of providing a correct prediction.

%\begin{quote}
%    for simple cases, it works really well. But for complicated ones, you need to make amendments. -- emrah
%\end{quote}

\begin{quote}
    ``When there's an advanced context or when the context is very specific, then it [the RAG system] doesn't seem to give a useful answer'' -- Karl, Case handler
\end{quote}

The filtering function was the first instance where the data scientists had to give up trying to solve their data requirement challenge by adding more data. Instead, they developed filtering as compensation for this limitation. This filtering function allowed the case handlers to provide the missing contextual information and determine which previous answers should inform the generated response. As a result, case handlers could input the relevant contextual data themselves for the model to generate an appropriate answer.

%There are too many execptions from the rules that our data will cover it all.
%By combining their professional experience with an understanding of the model's functionality, casehandlers can assess what information the model can and cannot reliably provide. One casehandler explains further:

%\begin{quote}
%   \hl{"I check a bunch of things outside the model's grasp that I need to construct a good answer." -- Case handler}
%\end{quote} The ability to evaluate the model's accuracy and determine which scenarios fall within or outside its capabilities improves with the casehandlers' domain expertise and understanding of machine learning principles.

\subsection{Engineering for (Ir)retrievable Data}
Despite the filtering function addressing some retrieval problems, other challenges remained. The nature of rules is that they require interpretation, and for certifying companies like Marcomp this means that every answer must comply with the prevailing interpretation norms among classification societies. These norms are established in joint forums among classification companies and regulatory authorities. Deviating from these norms could damage Marcomp's reputation and, in the worst case, they could lose their authority to issue certificates.

During one of our observations, a case handler received a question she recognized as having a strategic motive behind it. 
\begin{quote}
    ``The straight forward "correct" answer would favor the wishes of the captain here. However, in this context, we need to uphold the consensus on how this rule is interpreted and write a more strategic answer.'' -- Kat, Case handler
\end{quote} 
The case handler pointed out that even when the model creates text that looks flawless, it might fail to capture the key interpretation elements and potentially undermine established consensus in the industry. Case handlers preserve this established precedent in their responses. Senior case handlers serve a vital function in these scenarios. Their years of experience allow them to recognize complicated rules and understand how they should be interpreted according to international agreements.

Again, unable to adequately solve this data requirement, the data scientists had to compensate with a different approach. They ensured that the human case handler always checks every generated response before sending it, even for the simplest cases. While it would theoretically be possible for the RAG system to recognize interpretation norms from previous answers and generate correct responses, this was not practically possible with the existing database of 500,000 answers.

\subsection{Managing Expectations}

To identify the RAG system's correctness and find measures to improve it, like the filters, the data scientists first had to release the system and see how case handlers used it in real situations. The data scientists stressed that finding the key contextual elements and determining which ones were missing from previous answers completely depended on observing case handlers' reactions to early versions of the system. This approach helped them discover what data was missing from the database and what would be impossible to include. These insights guided their development of the filters.
\begin{quote}
``We know that the model [RAG system] will make errors, it is inevitable. So we have to think completely differently than normal, we have to think about the consequences when it is wrong, and set up safeguards for that. [...] We have to somehow get the system started even though it's not perfect, and then based on that we can initiate continuous improvement.'' -- Heimdal, AI Solution Engineer
\end{quote}

However, this approach put the data scientists in a difficult position of managing users' high expectations while needing to release an early, imperfect version of the model. While observing the case handlers, we noticed that they did not feel threatened by introducing the RAG system, despite initial comments that it could take over their job. After testing the system, they doubted it could match their performance due to the complex contexts and strategic elements involved in answering questions. Case handlers shared that their high expectations were unmet in interviews and observations. Many were initially impressed by their first experience with ChatGPT and hoped for the same level of impact again.
\begin{quote}
    ``I feel that this AI is kind of a toddler. ... I'm like a God to him'' -- Emil, Case handler
\end{quote}

Despite users' high expectations, the data scientists knew the system would make mistakes because data requirements were difficult to meet, especially in early versions. They made some efforts to manage the users' high expectations but eventually gave up. The hype surrounding LLMs seemed too strong and all-encompassing. They still chose to release it to users because they needed to discover perceived correctness of generated answers and ways to continuously improve it.

\begin{quote}
    ``I have given up trying to explain this [to the users], and just started implementing.'' -- Heimdal, AI Solution Engineer
\end{quote}

\section{Towards Requirements Engineering for RAG}

\begin{figure}
    \centering
    \includegraphics[width=1\linewidth]{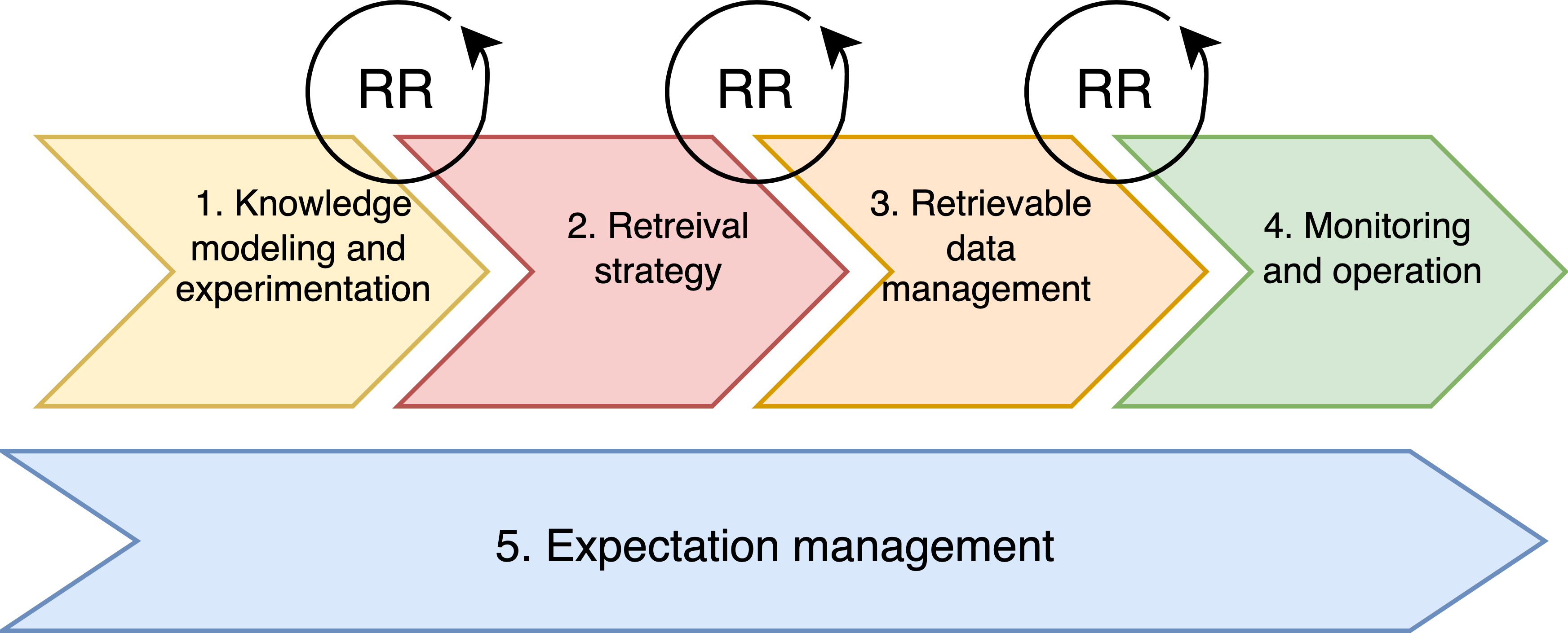}
    \caption{An iterative five-stage process model for eliciting "retrieval requirements" (RR). First, data scientists explore the available data in the knowledge base. Second, they define which parts should be retrievable as input to the LLM. Third, they work with users to assess the RAG system’s output and design filtering functions for tailored retrieval. Fourth, they monitor the live system to identify new "retrieval requirements". Fifth, they continuously manage user expectations as the system evolves. }
    \label{fig:rag process}
\end{figure}

In addressing our research question: \emph{how do developers elicit requirements for RAG-systems?}, we propose a five-stage iterative process that outlines how developers elicit "retrieval requirements" for RAG systems (figure \ref{fig:rag process}). We identify four sequential steps: 1) Knowledge modeling and experimentation, 2) Retrieval Strategy, 3) Retrievable data management, and 4) Monitoring and operation. With an underlying continuous activity of expectation management (5). 

This process shares commonalities with existing machine learning processes \cite{amershi_software_2019}, particularly regarding the (1) \textit{Knowledge modeling and experimentation} stage, where developers must evaluate what information exists in the organization's knowledge base and which potential LLMs might be suitable for implementation.

However, our findings reveal two distinctive stages: \emph{Retrieval strategy} and \emph{Retrievable data management}. The (2) \emph{retrieval strategy} stage enables data scientists and developers to define which data the LLM can access when generating predictions—in other words, what it can and cannot retrieve as part of the prompt. This phase involves searching and matching the incoming questions or problems with the existing knowledge base at runtime. This highlights a key difference between RAG and traditional ML: RAG allows flexible, runtime control \cite{barnett_seven_2024}, while traditional ML requires model retraining to handle input changes \cite{vogelsang_requirements_2019}.

The (3) \emph{Retrievable data management} stage addresses knowledge that can be effectively retrieved, providing case handlers with functions like retrieval filters or discarding generated content entirely. In this stage, developers create functions that empower users to address missing content, poorly ranked documents, and extraction failures as identified by \citet{barnett_seven_2024}. 

The final stage, (4) \emph{Monitoring and operation}, parallels the monitoring step described by \citet{amershi_software_2019}, emphasizing the need to evaluate system performance and determine if adjustments to the \emph{Retrieval strategy} are necessary due to changes in \emph{Retrievable data management}.

Throughout the process, (5) \emph{expectation management} is a foundational element. This emerges from uncertainty about user performance expectations and the RAG system's ability to generate "correct" answers. As shown in ML development research \citep{tanweer_why_2021, amershi_software_2019}, these systems heavily depend on user feedback and input to achieve satisfactory performance. Our findings demonstrate how data scientists elicited "retrieval requirements" through experiments and field studies, incrementally implementing additional filters to enhance the retrieval of previous answers, discovering necessary filter types through iterative implementation cycles.

\section{Conclusion \& Limitations}
Based on our preliminary findings, eliciting "Retrieval Requirements" is essential when developing RAG systems to ensure output correctness. This short paper presents an empirically based process for developing RAG systems and practical examples of how Retrieval Requirements can be identified. We plan to continue studying Marcomp's RAG system development to uncover additional aspects and refine our proposed development process. 

There are three major limitations to this study. First, as this is a case study, generalization was not the objective. Although our findings and the proposed process model for developing RAG systems appear logical and potentially applicable to other cases, a single case study cannot confirm their broader applicability. To achieve generalizability, a quantitative study would be preferable. Alternatively, conducting multiple case studies across different industrial contexts could strengthen the conclusions and reveal patterns across settings.

Second, construct validity poses a challenge due to the immature state of terminology surrounding AI, ML, and RAG systems among practitioners. Informants often use different terms for the same concept or the same term for different concepts, complicating the research process. This required us to frequently validate our interpretations of their language. To improve construct validity in this emerging topic, we urge researchers to develop empirically grounded definitions of the concepts studied. Addressing construct validity explicitly is essential when researching new areas such as requirements engineering for RAG systems.

Third, we have not yet measured the effects of the RAG system at Marcomp, meaning we cannot determine whether its development has been a success. In future work, we plan to study the system's impact on productivity and knowledge sharing among case handlers. We are particularly interested in whether the RAG system can handle most simpler questions automatically, thereby removing routine tasks from case handlers. At the same time, we aim to examine how this might affect junior case handlers, who typically rely on these simpler tasks for early career learning. Additionally, we want to investigate whether automating simple tasks changes the nature of questions juniors direct to seniors, potentially altering knowledge-sharing dynamics.

\section{Acknowledgments}
We thank the Norwegian Research Council for funding this research (grant number: 309631 \& 355691).

\end{document}